\documentclass[twocolumn,prl,floatfix]{revtex4}
\usepackage{graphicx}

\usepackage{epstopdf}
\usepackage{dcolumn}
\usepackage{amssymb,amsmath}
\usepackage{bm}

\begin{document}

\title{Hard Discs on the Hyperbolic Plane}
\author{Carl D. Modes and Randall D. Kamien}  \affiliation{Department of Physics and Astronomy, University of Pennsylvania, Philadelphia, PA 19104-6396, USA}

\date{\today}

\begin{abstract}
We examine a simple hard disc fluid with no long range interactions on the two dimensional space of constant negative Gaussian curvature, the hyperbolic plane.  This geometry provides a natural mechanism by which global crystalline order is frustrated, allowing us to construct a tractable model of disordered monodisperse hard discs.  We extend free area theory and the virial expansion to this regime, deriving the equation of state for the system, and compare its predictions with simulation near an isostatic packing in the curved space.
\end{abstract}

\maketitle

The precise connections between the virial equation of state, the crystalline transition, glass transitions, and random close packing remain elusive \cite{Torquato, dohard?}.  The systematic calculation of the equation of state is perturbative and analytic in volume fraction, $\phi$ \cite{landau}, and thus it is difficult to comprehend how it can, taken at face value, probe the nonanalytic, large volume fraction regime where crystallization occurs.  At the other extreme, crystal free energies are well modeled by Kirkwood's free volume theory \cite{Kirkwood}, especially near their maximum density.   How can we probe, in three-dimensions for instance, the crystalline phase transition in hard spheres   at $\phi_X\approx 0.5$ and the possible existence of a maximally-random jammed state near $\phi_{RCP}\approx 0.64$ \cite{Torquato,KamienLiu}?  
Focussing on two-dimensional systems as a simple
starting point is fraught with difficulties.  Unlike the situation in three dimensions where local tetrahedral and icosahedral packing competes with global fcc crystallization \cite{NelsonPRL}, 
it is problematic to prevent crystallization in simulations of monodisperse discs \cite{ohern}, as triangular close-packing is commensurate with crystalline order. Here we formulate the problem on a curved surface which serves to frustrate global crystalline order in a manner analogous to that in three dimensions.  In particular, we study hard discs on the hyperbolic plane, ${\mathbb{H}}^2$, at a curvature near a known regular tesselation.  This model is numerically tractable and we find the equation of state for hard discs via molecular dynamics.  We compare our results to free-area theory for the packing derived from the nearby tesselation.

Experimentally, one could consider crystalline order or packing problems on any of the many bicontinuous structures in lipid and diblock phaes \cite{Holyst} and present in the cell, such as the periodic cubic membrane in the mitochondria of \textit{Chaos carolinensis} \cite{bingo}.  These bicontinous structures, often close to minimal surfaces, have negative intrinsic curvature and the regular tesselations we will consider here should serve as a scaffolding to model transitions on these surfaces. 

  On a two-dimensional manifold with constant intrinsic curvature $K$, the area of a circle of radius $r$ is $a=A(r)=-2\pi K^{-1} \left[\cosh(\sqrt{-K}r) -1\right]$ (note that the $K\rightarrow 0$ limit recapitulates the Euclidean plane).  When we insert a disc of radius $r$ onto our manifold, it excludes an area of radius $2r$ to the other discs.  Thus the number of discs that cannot be inserted once the first disc is placed $A(2r)/A(r)\sim \exp\{\sqrt{-K}r \}$ grows exponentially with curvature and so we should expect that large, negative curvatures will lead to highly correlated disc packings due to the harsh entropic cost of this excluded area.  Equivalently, the kissing number, $n_{\rm kiss}$, the number of spheres which may simultaneously touch a central sphere, as a function of the curvature, $K$, is \cite{nelson}:
\begin{equation}
n_{\rm kiss} = \frac{2\pi}{\cos^{-1} \left( \frac{\cosh (2\sqrt{-K}r)}{1 + \cosh (2\sqrt{-K}r)} \right)}\underset{K\rightarrow -\infty}{\sim}\pi \exp^{\sqrt{-K}r}
\label{kn}
\end{equation}
To explore this systematically, we
study
the virial expansion for  the pressure $P$ in terms of the number density $\rho$, written
$
P=k_BT\sum_{n=1}^\infty B_n\rho^n
$
where the $B_1$ term arises from the usual ideal gas law.  While it may be tempting to assume $B_1=1$, the ideal gas on a curved manifold requires some thought;   the circumference of a circle,
$C(r)=2\pi  \sinh(\sqrt{-K}r)/\sqrt{-K}$, grows as fast as $A(r)$ making the infinite area, thermodynamic limit problematic.

The textbook approach to the ideal gas law begins with
particle-in-a-box eigenenergies and takes the large area limit to calculate the 
partition function for a single particle, $Z_1=V/\lambda_T^d$ where $\lambda_T=\sqrt{2\pi\hbar^2/(m T)}$ is
the thermal wavelength.   Fortuitously, the spectrum of the Laplacian on a general Riemann surface
has been well-studied.  McKean and Singer \cite{MS} developed the Weyl expansion \cite{Weyl} for the partition function of a single particle on a periodic two-dimensional domain $D$:
\begin{eqnarray}
Z_1(T) 
&=& \frac{1}{\lambda_T^2}\int_D dA \pm \frac{1}{4\lambda_T}\int_{\partial D} d\ell +\frac{1}{12\pi}\int_D K dA\nonumber\\
&&\quad  - \frac{1}{6\pi}\int_{\partial D} H d\ell +{\cal O}(\lambda_T)
\end{eqnarray}
where the sign of the first correction depends on Neumann (+) or Dirichlet (-) boundary conditions and $H$ is the mean curvature. Now we see that the circumference directly affects $B_1$ and the only way to avoid the problem with the boundary is to consider periodic boundary conditions, {\sl i.e.} no boundary.
We note that for hard discs the virial expansion only includes local interactions among clusters of particles.  As long as the clusters in question do not wrap around the periodic space, we ignore the periodicity and perform the cluster integrals on ${\mathbb{H}}^2$.  In order to ensure that this is possible, we may continue to increase the area available to a cluster by constructing ever higher genus manifolds \cite{tarjus} and, thus, we can neglect ``winding'' of the discs around the periods.  For manifolds with constant $K<0$, $Z_1=A[\lambda_T^{-2} +K/(12\pi)]>0$ when 
$\lambda_T^2<-12\pi/K$.  In other words, we are forced to work in the regime where the curvature of the manifold does not probe the quantum regime of the gas.  In this regime, the pressure
$P=Nk_{B}T d\ln Z/dA$ is independent of the proportionality constant.  

To calculate the virial coefficients, we employ the Poincar\'{e} disc model of $\mathbb{H}^2$, popularized by Escher \cite{escher}, which conformally maps the hyperbolic plane  with curvature $K$ to the disc of radius $\sqrt{-K^{-1}}$ on $\mathbb{R}^2$.  Using the co\"ordinates $(x,y)$, the metric is \cite{tarjus}
\begin{equation}\label{metric}
ds^2 = \frac{4\left(dx^2+dy^2\right)}{\left[1 +K\left(x^2+y^2\right)\right]^{2}}
\end{equation}
where $x^2+y^2\le -1/K$. 
Geodesics in this model are arcs of circles which intersect the bounding circle normally;  circles in the hyperbolic plane remain circles in this model, but their coordinate centers and radii vary.
The radius of curvature sets a lengthscale and, consequently, any system on $\mathbb{H}^2$ is manifestly not scale invariant.  

We take advantage of this lack of scale invariance to probe higher curvatures not by changing the ambient space, but simply by considering larger and larger discs.  We consider discs of radius $r\sqrt{-K}$.  Without loss of generality we fix the Gaussian curvature to be $K = -1$.  Dimensional analysis may always be employed to insert the appropriate factors of $K$. 
As in flat space, the 
second Virial coefficient, $B_2(r)$, is given by half the excluded area of a single disc:
$B_2(r) = \pi\left[\cosh(2r) - 1\right]$.
As $r\rightarrow 0$, $B_2\sim 2\pi r^2$, we recover the $d=2$ Euclidean result:
$B_2(r) = (4\pi)^\frac{d}{2} r^d/ \Gamma(\frac{d}{2}) \, d.$
Though it may been tempting to associate the higher connectivity at larger curvatures with higher dimensions, we know of no natural identification.

\begin{figure}[t]
\centering
\includegraphics[width=7cm]{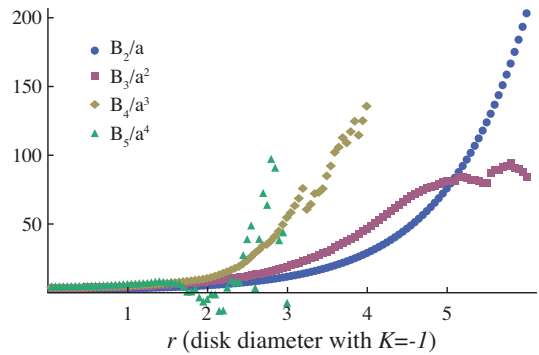}
\caption[B3B4B5plot]{(Color online).  The second, third, fourth and fifth virial coefficients in appropriate units of the area, $a$, as functions of disc size (for $K =-1$) $r$. \label{B3B4B5plot}}
\end{figure}

We turn to numerical integration to evaluate the higher virial coeffecients.  In addition, as is traditional in flat space \cite{lubanbaram}, the higher coefficients $B_n$ are reported in units of $a^{n-1}=\left[A(r)\right]^{n-1}$ to suppress size effects and generate the expansion for
$P/(\rho k_{B}T)$ in the area fraction, $\phi=\rho a$:
\begin{equation}
P=\rho k_{B} T\left[1 + \sum_{n=2}^\infty \frac{B_n}{a^{n-1}}\phi^{n-1}\right]
\end{equation}  \label{phivirial}
We have checked the $r\rightarrow 0$ limit in all cases, and happily find agreement with the known Euclidean results.
We plot $B_2/a$, $B_3/a^2$, $B_4/a^3$, and $B_5/a^4$ in Figure \ref{B3B4B5plot} as a function of $r$.    To the accuracy of our calculation, $B_3$ and $B_4$ are positive,  while $B_5$ may have a zero crossing.  There is evidence that, in higher dimensional flat spaces, the virial coefficients alternate in sign \cite{lubanbaram, virialD} and hence the leading singularity that controls the radius of convergence of the expansion sits on the negative real axis \cite{virialD}.  The location of this singularity, and hence the range of applicability of the virial expansion itself, remains an open question in $D = 2,3$. Whether or not $B_n(r)$ crosses zero for the negatively curved system at any value of $r$ is the subject of future work.  

Since our confidence in the reliability of the virial expansion wanes with ever larger curvature, we are forced to look at highly correlated structures on $\mathbb{H}^2$.  To start we consider regular packings on the hyperbolic plane.  We begin with a regular lattice of Voronoi cells and 
consider a tesselation of $F$ $p$-gons, with $q$ meeting at each vertex. The 
total number of vertices is $pF/q$, the total number of edges is $pF/2$, and the total number of
faces is $F$.  The Euler character $\chi$ of the surface constrains these three numbers and it follows
that 
\begin{equation}\label{euler}
\chi=V-E+F = pF\left[\frac{1}{ q}  + \frac{1}{p}- \frac{1}{2}\right]
\end{equation}
For the flat periodic plane ({\sl i.e.} the torus) $\chi=0$, so $p^{-1}+q^{-1}=1/2$ and we find the three tilings represented by the Schl\"afi symbol $\{p,q\}=\{6,3\}$, $\{3,6\}$, and $\{4,4\}$, hexagons, triangles, and squares, respectively.  The spherical topology is the only one with positive 
$\chi=2$, admitting the five Platonic solids $\{p,q\}=\{3,3\}, \{3,4\},\{3,5\},\{4,3\}$, and $\{5,3\}$.  Finally, we turn to the pertinent geometries with negative Gaussian curvature, for
which $\chi<0$ and $p^{-1}+q^{-1} < \frac{1}{2}$.  In this case there are an infinite number of integral pairs $\{p,q\}$ and an infinite number of regular tesselations.  Triangular packing around each vertex, $q=3$, corresponds to close-packing.   We focus instead on isostatic packings, those for which each
disc has $p=4$ nearest neighbors.  Recall that an isostatic packing is one for which the number of force balance equations is equal to the number of degrees of freedom.  For $N$ frictionless particles in $d$ dimensions with coordination $z$,
there are $dN$ degrees of freedom and $Nz/2$ forces to balance: isostaticity requires $z=2d$ \cite{maxwell}.

We find the area fraction of $\{p,q\}$ through application of the Gauss Bonnet Theorem and hyperbolic trigonometry.  When there are $F$ identical polygons, the area of each can be found from the Euler character $\chi$: $F A_{\rm p-gon} = \int dA = -\int K dA = -2\pi \chi$,
where we have used $K=-1$.  The radius of the incircle can be determined via the dual law of hyperbolic cosines \cite{coxeter},  $\cosh(r) = \cos(\pi/q)/\sin(\pi/p)$.   We find the area fraction for the $\{p,q\}$ tesselation:
\begin{equation}
\phi_{\{p,q\}}= \frac{\frac{\cos(\pi/q)}{\sin(\pi/p)} - 1}{p\left(\frac{1}{2} - \frac{1}{q} -\frac{1}{p}\right)}
\label{pqformula}
\end{equation}

We recover the appropriate packing fractions, $\pi/3\sqrt{3}$, $\pi/4$, and $\pi/2\sqrt{3}$, for the Euclidean tilings, $\{3,6\}$, $\{4,4\}$, and $\{6,3\}$, respectively.  As $p\rightarrow\infty$ for close-packed $q=3$ tesselations, we find $\phi_{\rm max} = 0.9549$, the known packing fraction of the best packing on the hyperbolic plane at any curvature \cite{twelve}. The lowest packing fraction for an isostatic configuration ($p=4$) is $\phi_{\rm iso}=\phi_{\{4,\infty\}}=\sqrt{2}-1\approx 0.4142$, an area fraction far below that for the Euclidean square lattice $\pi/4\approx 0.785$.  Also of note is the broadening of the range in $\phi$ that supports stable configurations: $\phi_{\rm max}-\phi_{\rm iso}\approx 0.54$ on $\mathbb{H}^2$, as compared to a range of roughly $0.13$ in flat space.  This is consistent with our finding that the virial expansion breaks down at lower area fractions on $\mathbb{H}^2$.  Indeed, the $\{4,\infty\}$ tesselation requires $\cosh(r^*)=\sqrt{2}$ so that $B_2(r^*)/A(r^*) = \frac{1}{2}(2)/(\sqrt{2}-1)=1/\phi_{\{4,\infty\}}$.  Though possibly a mathematical profundity, the fact that $B_2(r^*)\phi_{\{4,\infty\}}/A(r^*)=B_1$ means that the first correction to the ideal gas is precisely equal to the non-interacting result 
and strengthens the case that we must study the large $\phi$ regime.

\begin{figure}[t]
\centering
\includegraphics[width=7cm]{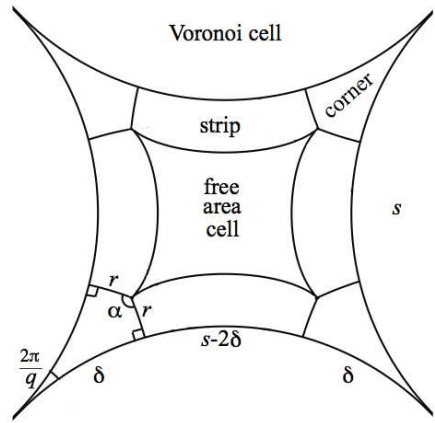}
\caption[pqFAT]{Partition of a $\{p,q\}$ Voronoi cell into its free area cell, $p$ corners, and $p$ strips.  Note that the boundaries of the free area cell are not geodesics.  \label{pqFAT}}
\end{figure}

The presence of these large corrections, even at low area fraction, necessitates an approach to the high-density regime.  We turn to free-area theory, known to be exact near close-packing \cite{Salzburg,Torqu}.  In flat space, scale invariance ensures that the shape of the free-area region, available to each disc's center of mass, has precisely the same
shape as the corresponding Voronoi cell.  However, the rigidity of $\mathbb{H}^2$ forces us to consider
how the shape of the cell changes as the size of the discs change.  In Figure \ref{pqFAT} we show how
we break up the Voronoi cell to calculate the resulting free area.  

\begin{figure}[t]
\centering
\includegraphics[width=8.5cm]{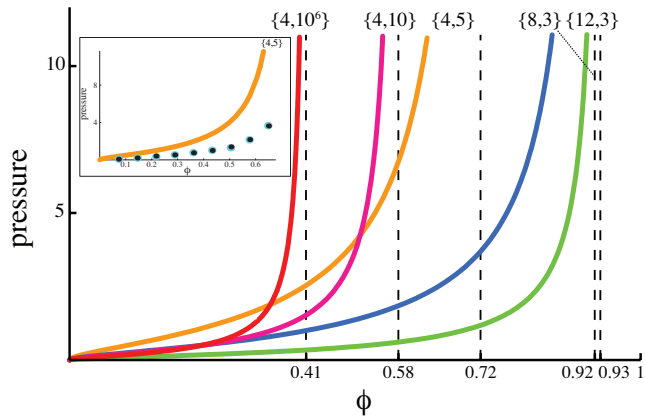}
\caption[disccell]{(Color online).  The equation of state for free area theory for a range of $\{p,q\}$ tesselations.  Note that the pressure curves diverge at their maximum packing fractions and that they differ even at low volume fractions, $\phi$. The vertical dashed lines indicate the maximum volume fraction for the tesselations listed on top of the figure.   The inset compares our numerical results with the $\{4,5\}$ free-area theory at larger values of the pressure.  The larger, light blue circles are the result of our molecular dynamics simulation for curvature just below the $\{4,5\}$ tesselation and the smaller, black circles are for curvature just above $\{4,5\}$. The number of particles ranges from 1 to 9.  \label{disccell}}
\end{figure}

The free-area can be found by starting with $A_{\text{p-gon}}$ and subtracting the area of each of the $p$ strips, $A_{\rm strip}=(s - 2\delta)\sinh(r)$ \cite{strips} and each of the $p$ corners $A_{\rm corner}=\pi-\alpha-2\pi/q$.  The hyperbolic law of cosines \cite{coxeter} allows us to find $\cosh \delta =\cos(\alpha/2)\csc(\pi/q)$,  $\cosh s=[\cos(2\pi/p) + \cos^2(\pi/q)]\csc^2(\pi/q)$, and $\sin(\alpha/2)=\cos(\pi/q)/\cosh r$.  We can relate $r$ to $\phi$, $p$, and $q$ and write $A_{\rm free}$ in terms of $p$, $q$, and $\phi$: setting $\tau = 1 - \frac{\chi \phi}{F}$ and $\tau_{pq}=1-\frac{\chi\phi_{\{p,q\}}}{F}$, we have
\begin{eqnarray} A_{\text{free}}  &= & 2p \sin^{-1} \left[\sin\left(\frac{\pi}{p}\right) \frac{\tau_{pq}}{\tau} \right]- 2 \pi\nonumber\\ &&- 2p \sqrt{\tau^{2} - 1}
\cosh^{-1} \left[ \frac{\cos(\frac{\pi}{p})}{\sin(\frac{\pi}{q})} \right] \nonumber \\
& & +2p \sqrt{\tau^{2} - 1} \cosh^{-1} \left[ \frac{\sqrt{1-\frac{\tau_{pq}^2}{\tau^{2}}\sin^2(\frac{\pi}{p})}}{\sin(\frac{\pi}{q})} \right] \end{eqnarray}
Unfortunately, this expression is cumbersome to the point of uselessness, so
we instead turn to Figure \ref{disccell}, where we plot the free-area equations of state for various $\{p,q\}$ tesselations using 
$P_{FA} = -T a^{-1} \phi^{2} \frac{d}{d\phi} \ln(A_{\rm free})$.  Note that $a$ is set by $A_{\text{p-gon}} \phi = a$.

Finally, we turn to disordered packings.  We employed constant energy molecular dynamics on a periodic hyperbolic octagon with sides identified to give the topology of the $2$-torus \cite{tarjus}.  Note that we cannot vary $\phi$ by changing the disc size, as we must in free area theory, since that is, by the absence of scale invariance, equivalent to a change in the background curvature.  Instead, we change $\phi$ by changing the number of discs in the periodic octagon.    We could choose a larger, higher genus periodic cell on which to do our simulation in which case we could probe a larger number of area fractions, filling in the data points in Figure \ref{disccell}, though this is computationally intensive.  Although the simulation and free-area theory share a common endpoint, the pressure in the latter samples ever-changing curvatures on its way to close packing. 
We chose to study curvatures $r=0.53$ and $r=0.531$, just {\sl below} and just {\sl above} that for the $\{4,5\}$ tesselation, $r_{\{4,5\}}=0.5306$.  At these curvatures no regular tesselation is allowed, and so non-crystalline arrangements give the best packings \cite{radin}.  Thus we expect to probe disordered configurations even at high densities with this choice.  Indeed, the simulation suggests that the disordered system attains higher pressures less rapidly than the nearby, ordered, isostatic configurations, particularly at larger values of $\phi$ (Figure \ref{disccell}).  The $r\rightarrow 0$ limit should be a way
to study these disordered packings on the Euclidean plane.  Fortuitously, roughly 10 discs fill the periodic octagon, the smallest periodic region on $\mathbb{H}^2$, and thus we could quickly approach a dense packing.  We require many more discs to study the Euclidean limit; this work is in progress.

We studied a hard disc gas through simulation, at curvatures close to that for a regular tesselation and compared with free-area theory.  Thus we have a model which approaches disordered packings of monodisperse discs for which we can understand the equation of state.  Note that our approach allows us to ponder the structure of the virial expansion:  in $d$ dimensions, the $2d^{\rm th}$ virial coefficient, $B_{2d}$ is the required to probe isostaticity, while crystallization requires the $n_{\rm kiss}^{\rm th}$
virial coefficient.  Though increasing the dimension $d$ pushes $n_{\rm kiss}$ ever larger \cite{Exponential}, the study of 
disordered packings forces us to calculate ever more virial coefficients.  The hyperbolic plane sidesteps
this problem by keeping isostaticity at $p=4$.   Further work will focus on other tesselations, in particular, the Euclidean limit, and on mapping the results and predictions of $\mathbb{H}^2$ onto the bicontinuous structures present in biological systems.

It is a pleasure to acknowledge discussions with N. Clisby, C. Epstein, A. Liu, M. L\'{o}pez de Haro, B. McCoy, C.D. Santangelo, and D. Vernon.  This work was supported in part by NSF Grant DMR05-47230, a gift from L.J. Bernstein, and a gift from H.H. Coburn.  RDK appreciates the hospitality of the Aspen Center for Physics where some of this work was done.

\end{document}